\documentclass[prd,twocolumn,floatfix,english,letterpaper]{revtex4}

\usepackage{amssymb}
\usepackage[dvips]{graphicx}

\usepackage[T1]{fontenc}
\usepackage[latin1]{inputenc}

\makeatletter

\providecommand{\tabularnewline}{\\}

\usepackage{babel}
\makeatother
\begin{document}

\title{Excited Upsilon Radiative Decays}

\author{Randy Lewis}
\affiliation{Department of Physics and Astronomy, York University,
Toronto, Ontario, Canada M3J 1P3}

\author{R. M. Woloshyn}
\affiliation{TRIUMF, 4004 Wesbrook Mall, Vancouver,
British Columbia, Canada V6T 2A3}

\begin{abstract}
Bottomonium S-wave states were studied using lattice NRQCD. Masses of ground and
excited states were calculated using multiexponential fitting to a set of
correlation functions constructed using both local and wavefunction-smeared
operators. Three-point functions for M1 transitions between vector and
pseudoscalar states were computed. Robust signals for transitions involving
the first two excited states were obtained. The qualitative features of the 
transition matrix elements are in agreement with expectations. The calculated 
values of matrix elements for $\Upsilon(2S)$ and $\Upsilon(3S)$ decay are 
considerably larger than values inferred from measured decay widths.

\end{abstract}

\maketitle

\section{\label{sec_intro}INTRODUCTION}

The bottomonium $\Upsilon$ was discovered in 1977 \cite{herb77,innes77} and, 
remarkably, it took 30 years before its pseudoscalar partner $\eta_{b}$ 
was observed\cite{babar08,babar09}.
The measurement of the branching ratio for radiative decay of $\Upsilon(2S)$
and $\Upsilon(3S)$ to $\eta_{b}$ presents an opportunity to test
calculational methods where the decay amplitude depends entirely on small
effects: spin-dependent interactions, recoil and relativistic corrections.
In this paper we present a first pass at the calculation of the excited
Upsilon radiative decays using lattice NRQCD\cite{latnrqcd}.

To understand the challenge of excited Upsilon decays it is useful
to consider first the amplitude for the magnetic dipole (M1) transition
between vector and pseudoscalar states in the nonrelativistic quark
model\cite{godf01,bram06}
\begin{equation}
\mathcal{M}(nS\rightarrow n'S)=\intop_{0}^{\infty}R_{n'}(r)R_{n}(r)j_{0}(qr/2)r^{2}dr
\label{eq_overlapint}
\end{equation}
where $R_{n}(r)$ is the radial wavefunction for the S-wave state
with principal quantum number $n$ and $q$ is the photon momentum.
In a transition between states with the same principal quantum numbers
$n'=n,$ for example ground state to ground state, the radial wavefunctions
of the vector and pseudoscalar mesons are very similar. The overlap
integral is close to 1. However, when $n'\neq n$ the wavefunctions
are orthogonal in the extreme nonrelativistic limit. For these so-called
hindered transitions the amplitude is highly suppressed and depends
on the interplay of small effects coming from spin-dependent interactions,
effects of recoil and relativistic corrections\cite{godf01,bram06,ebert03}.

The use of lattice QCD methods to calculate the amplitude for vector
meson radiative decays was suggested long ago \cite{rmw86,cris92}. 
Recently, interest in this application of lattice QCD has been revived
and charmonium has been studied in detail \cite{dudek06,dudek09,chen11}. 
A number of different ground-state to ground-state transition amplitudes 
have been calculated involving not just S-wave but 
also P-wave states \cite{dudek06,chen11}. In Ref. \cite{dudek09}
radiative decays of
excited charmonium states were also considered. Excited states appear
as nonleading contributions to lattice QCD meson correlation functions.
This, combined with the suppression of the hindered M1 amplitude,
makes it a challenge to achieve good statistical accuracy for these
decays (see Table III in Ref. \cite{dudek09}). 

The application of lattice QCD to excited states is now a very active
research area. A primary goal of this work is to see how well we can
extract the excited state signal buried under the dominant ground-state
contribution. One way to deal with this problem is to use a
variational method \cite{luscher90,michael85}
with an appropriate (and large) set of basis operators.
An alternative which has been applied successfully to the calculation
of the spectrum of bottomonium is to use constrained multiexponential
fitting\cite{constrainedfit} to get the subdominant contributions. This method 
can work well if the lattice simulation data have high statistical precision 
(see, for example, Ref. \cite{davies10}). 
The gauge field ensemble used in this study is quite small, only 192 configurations,
but we reduce the statistical fluctuations by using multiple time sources
per configuration and by employing a spatial wall source. This allows
the extraction of robust signals for the 2S and 3S excited states. 

Section \ref{sec_lat} outlines the lattice QCD simulation. The gauge 
field configurations come from a 2+1 flavor dynamical simulation and were
provided by the PACS-CS Collaboration\cite{pacscs09}. 
The b quarks are described using a standard
$\mathcal{O}(v^{4})$ lattice NRQCD action \cite{latnrqcd,davies94,gray}
 with Landau link tadpole
improvement. Two-point correlation functions of pseudoscalar and vector
operators are discussed in Sec.~\ref{sec_twopnt}. Two-point function fit 
parameters, simulation energies and overlap coefficients were obtained, and these
are used unchanged in subsequent three-point function fits. As a check
of the calculation the simulation energies for the lowest three states
from our multiexponential fits were compared to the variational analysis
that could be done with our limited basis set. Good consistency was
obtained. As well, the overlap coefficients for the lowest three Upsilon
states were used to estimate leptonic decay widths with reasonable
agreement with experiment. Three-point functions and transition matrix
element results are discussed in Sec.~\ref{sec_threepnt}. A summary is 
given in Sec.~\ref{sec_summ}.

\section{\label{sec_lat}Lattice simulation}

The lattice gauge field configurations were generated and made available
by the PACS-CS Collaboration\cite{pacscs09}. The 2+1 flavor dynamical simulation
used the Iwasaki gauge field action ($\beta$=1.90) and the clover-Wilson
fermion action. We use only the ensemble which is nearest the light-quark
physical point with pion mass 156MeV. The light and strange hopping
parameters are $\kappa_{u/d}=0.13781$ and $\kappa_{s}=0.13640$.
The number of lattice points is 32$^{3}\times$64 and the lattice
spacing $a=0.0907(14)$fm was determined by PACS-CS\cite{pacscs09} (along with
the light and strange hopping parameters) using the pion, kaon and $\Omega^-$
baryon masses as input, \emph{i.e.}, heavy-hadron input was not used in setting
the scale. The number of gauge
field configurations used was 192 (out of 198 available). For the
tadpole-improved NRQCD calculation the average link in Landau gauge
was used. The numerical value was estimated to be 0.8463.

The heavy quark is described using lattice NRQCD\cite{latnrqcd,davies94,gray}. 
The exact form of the Hamiltonian may be found in the Appendix of 
Ref. \cite{lewis09}. 
Terms up to $\mathcal{O}(v^{4})$ are kept in the nonrelativistic expansion 
which in the notation of \cite{lewis09} means $c_i=1$ for $i\leq6$ and $c_i=0$ 
for $i\geq7$.
The b-quark bare mass was determined
by fitting the kinetic mass to the observed mass of $\eta_{b}$ and
takes the value 1.945 in lattice units. The stability parameter $n$
appearing in the Hamiltonian was taken to be 4 in line with Ref. 
\cite{davies10}.

The simplest operators to use to describe the pseudoscalar and vector
states are the local ones which take the form $O(x)=\chi(x)\Gamma\psi(x)$
where $\psi(x)$ and $\chi(x)$ are nonrelativistic quark and antiquark
fields with $\Gamma$ equal 1 ($\sigma$) for pseudoscalar (vector)
mesons. To calculate ground-state properties a smearing of the local operators,
such as Jacobi smearing\cite{gusken89}, is often used to damp out the high-energy
modes created by local operators. However, for investigating excited
states it is more advantageous to include operators which suppress
the ground state. For this, wavefunction smearing\cite{davies94} is useful. 
The smeared operator takes the form $O(x)=\sum_{y}\chi(x)\Gamma\phi(x-y)\psi(y)$
where an effective smearing function was found to be\cite{davies10}
\begin{equation}
\phi(r)=(1-r/(2a_{0}))e^{-r/(2a_{0})}
\end{equation}
 which has the profile of the Coulomb S-wave first-excited state wavefunction. 
The parameter $a_{0}$ was taken to be 1.4(lattice units).
In addition to the smeared operator a doubly-smeared operator where
the wavefunction smearing was applied to both quark and antiquark
fields was included. Our complete set of meson operators consisted
of three types: local (l), smeared (s) and doubly-smeared (d). In
order to use the nonlocal smeared operators without gauge links connecting
the quark and antiquark the gauge field configurations were fixed
to Coulomb gauge.

\section{\label{sec_twopnt}Two-point functions}

\begin{figure}
\scalebox{0.50}{\includegraphics*{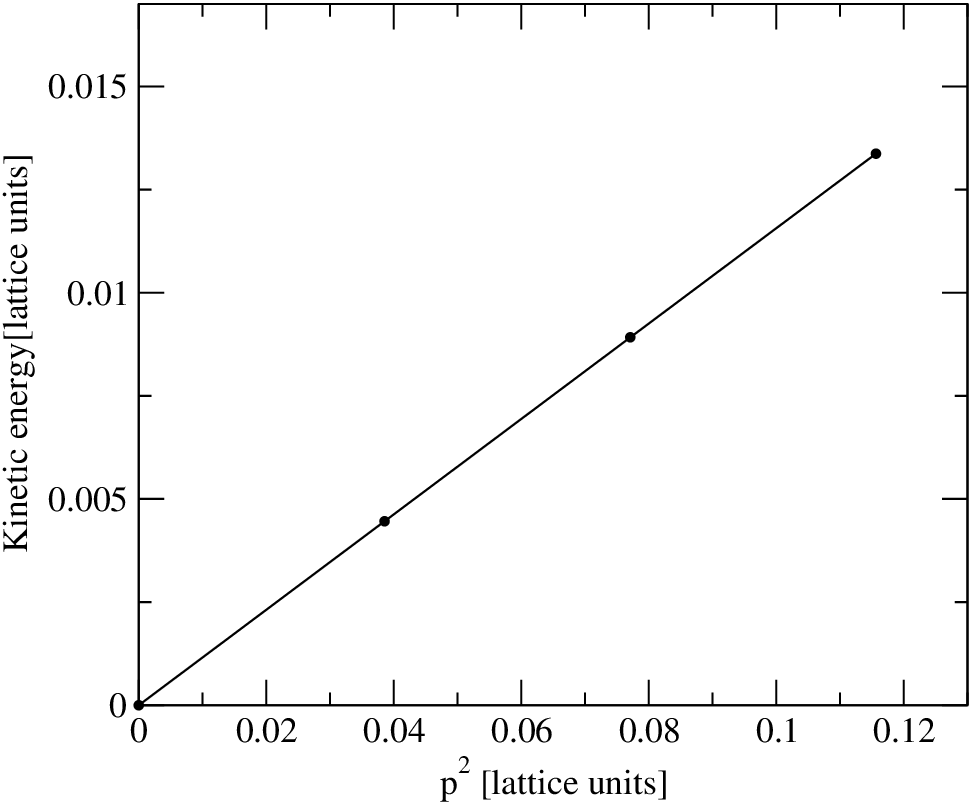}}
\caption{Kinetic energy of the pseudoscalar meson at b-quark bare mass $Ma =
1.945$ versus momentum squared. The line is a fit with Eq. (\ref{kinetic}). The
${\chi}^2/d.o.f.$ is 0.2.}
\label{kelat}
\end{figure}

In NRQCD the correlators of the meson operators do not give the hadron
mass directly. The simulation energy extracted from the zero-momentum
correlation function must be combined with the renormalized quark
mass and energy shift to get the meson mass. Alternatively the kinetic
energy can be used to determine the meson mass. This is the method
used here for tuning the b-quark mass to reproduce the mass of $\eta_{b}$.
Correlators of the local pseudoscalar meson operator projected onto
different values of momentum were calculated. Using the relation
\begin{equation}
E(p)-E(0)=\left(p^{2}+M_{0}^{2}\right)^{1/2}-M_{0}
\label{kinetic}
\end{equation}
the hadron mass $M_{0}$ can determined. Using this method we arrive
at a bare quark mass value of 1.945(4). The quoted error in this 
value reflects the uncertainty in the fit determining the kinetic
mass. As well, there is a 1.5\% uncertainty due to the uncertainty of
the lattice spacing determination. The kinetic energy and the fit
that determines $M_{0}$ at our nominal bare mass are shown in Fig.~\ref{kelat}.
The momenta used were (0,0,0), (1,0,0), (1,1,0) and (1,1,1) in units
of $2\pi/La$ where $La$ is the spatial extent of that lattice.

Using the determined value of the bare b-quark mass two-point correlation
functions (including cross correlators) of the three operator types
l,s,d were calculated for pseudoscalar and vector channels. Our lattice
has 64 time sites but it is not useful to construct correlators over
the entire time extent. The correlators are limited to maximum time
separation $t-t_{s}$ of 27. Since the maximum time separation is
considerably smaller than the lattice time extent and the nonrelativistic
propagators depend only on the gauge field links on time slices
between source and sink it is very effective to use multiple time
sources on each gauge configuration. Sixteen sources, uniformly distributed
in time, were used in this calculation. To further reduce statistical
fluctuations a random wall source (see, for example, \cite{davies07}) was used.
In addition to zero-momentum pseudoscalar and vector meson correlators,
pseudoscalar correlators with momentum corresponding to (1,0,0) and
(2,0,0) were calculated. These are needed for the analysis of the
three-point functions. 

\begin{figure}
\scalebox{0.55}{\includegraphics*{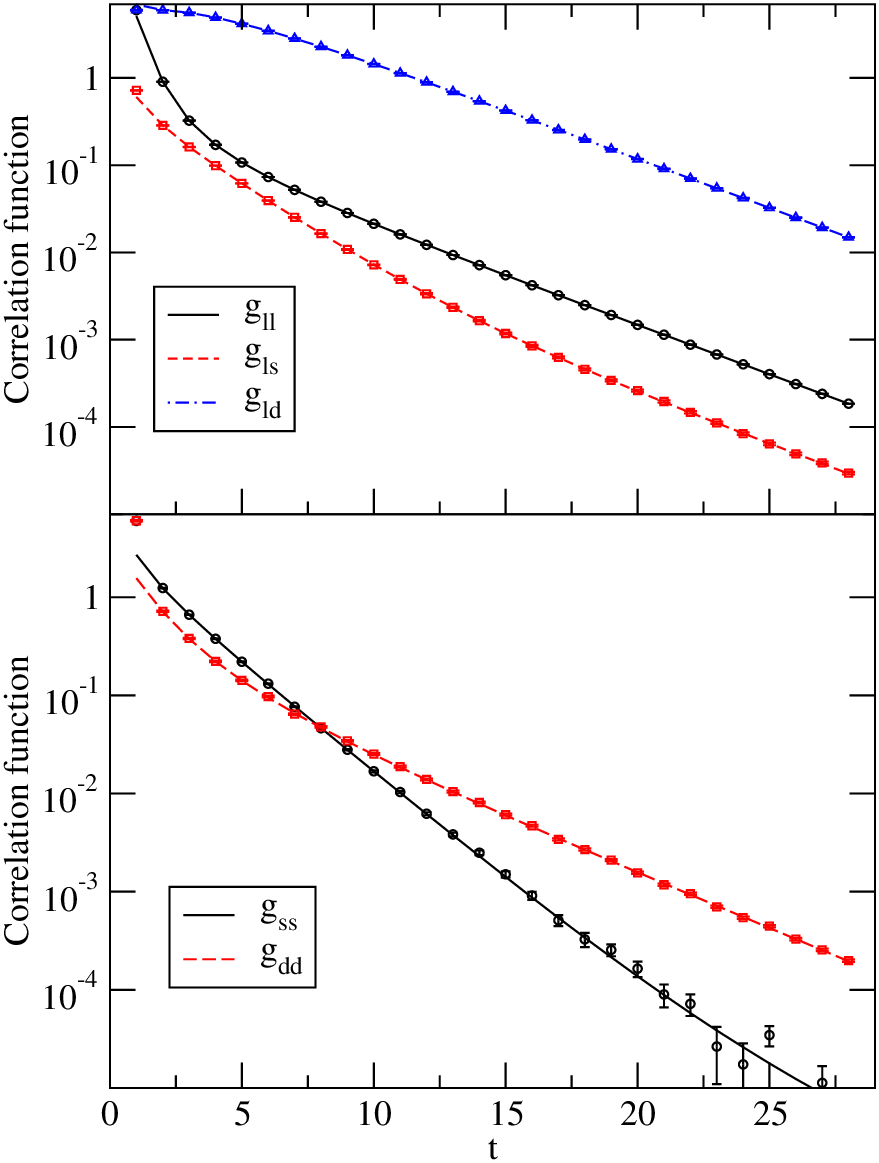}}
\caption{Zero-momentum vector correlation functions for different operator
combinations. Symbols are simulation values and lines are the result of a
fit with ten exponential terms. Except for some points with the ss
operator combination, statistical errors are smaller than the symbols.}
\label{twopntcf}
\end{figure}

Using subscripts $o$ and $o'$ ($o,o'=\{ l,s,d\})$ to denote source
and sink operators, the correlation functions $g_{oo'}(t)=\left<O_{o'}(t)O_{o}^{\dagger}(t_{s})\right>$(fixed
source time $t_{s}$) are fit with $N$ time-dependent exponentials
\begin{equation}
g_{oo'}(t)=\sum_{n=1}^{N}c_{o'}(n)c_{o}(n)e^{-E_{n}(t-t_{s})}.
\end{equation}
The constrained multiexponential fitting method\cite{constrainedfit,davies10} was used. 
All time points (except the source) were included. Using only loose constraints
fits are very stable even with a large number of exponential terms.
Figure \ref{twopntcf} shows a representative sample of correlation functions 
and fits, in this case for the zero-momentum vector channel with a simultaneous 
fit using 10 terms.

\begin{figure}
\scalebox{0.50}{\includegraphics*{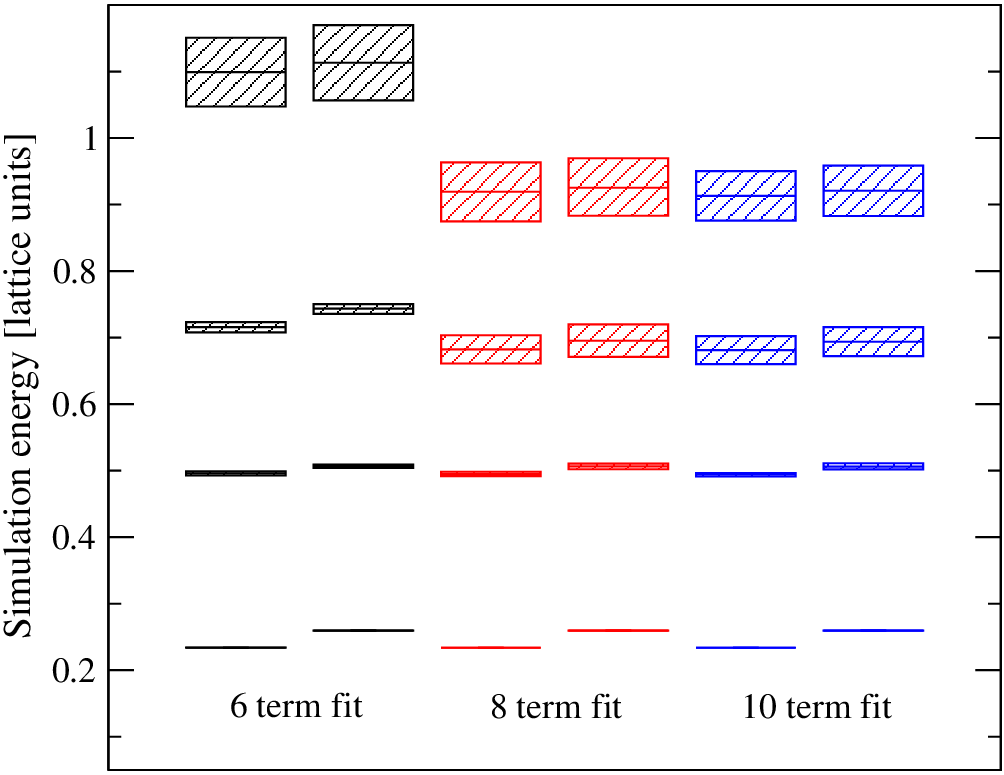}}
\caption{The simulation energies of the four lowest pseudoscalar and vector states from a
multiexponential fit to the two-point functions with 6, 8, and 10 terms.}
\label{Spectrum}
\end{figure}

\begin{figure}
\scalebox{0.50}{\includegraphics*{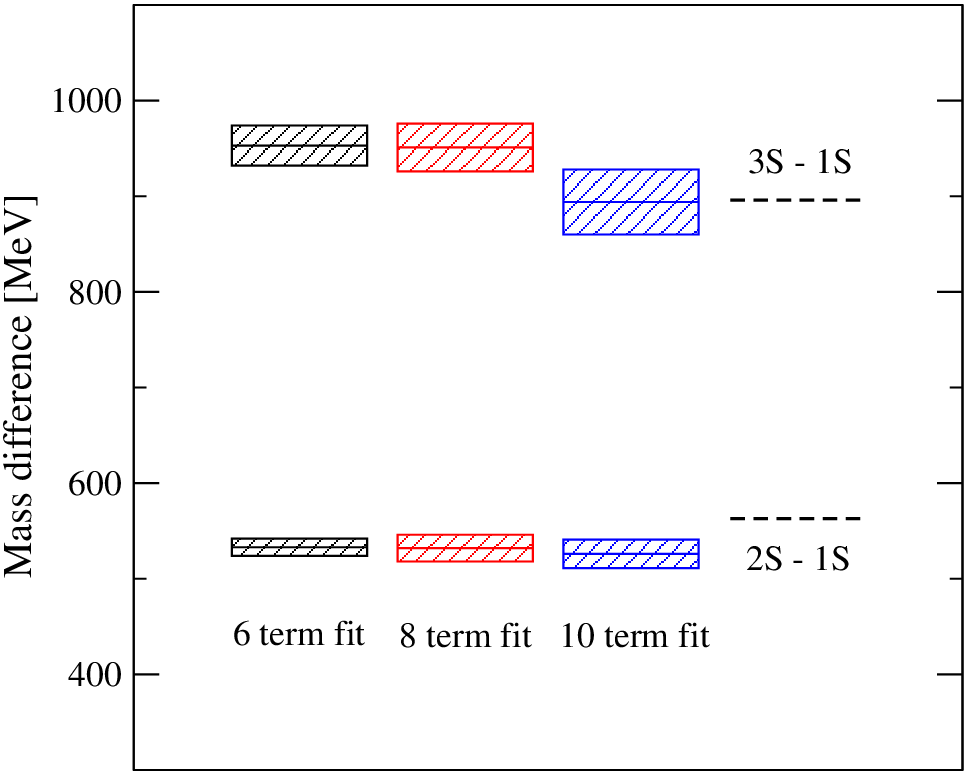}}
\caption{Mass difference between the Upsilon ground state and the first two
excited states using results of 6, 8 and 10 term multiexponential fits. The dashed lines
show experimental values using data from \cite{pdg10}.}
\label{massdiff}
\end{figure}

The lowest four simulation energies for zero-momentum pseudoscalar and
vector channels are shown in Fig.~\ref{Spectrum} for fits with 
6, 8 and 10 terms.
The lowest three states, which are of interest for our three-point
function calculation, are quite robust. Differences of the zero-momentum
simulation energies are just mass differences and these are shown
for the lowest Upsilon states in Fig.~\ref{massdiff}. The results 
are in reasonable accord with experimental values\cite{pdg10}.

The mass difference between $\Upsilon$ and $\eta_{b}$ was also calculated.
For the ground states, our result is 56(1)MeV where the error is dominated
by the uncertainty in the lattice spacing. This value is essentially
independent of which fit is used and consistent with values obtained
by others \cite{gray,meinel09} using the $\mathcal{O}(v^{4})$ lattice NRQCD action. It
is somewhat smaller than the PDG average\cite{pdg10} 69.8$\pm$2.8MeV of the
experimentally observed values\cite{babar08,babar09,cleo10}.
It is a common feature for lattice simulations of heavy quark systems 
to underestimate the spin splitting and some issues have been discussed in the
context of lattice NRQCD in recent studies\cite{meinel10,hamm11}. 

A popular way to determine excited state energies is the variational
method\cite{luscher90,michael85} (for an extensive recent discussion 
see \cite{bloss09}). 
The correlator matrix is diagonalized at each time and the
time-dependent eigenvalues then give an optimal estimate of the time
evolution of individual states. The evolution is calculated with respect
to some reference time $t_{0}>t_{s}$ which should be chosen large
enough so that the number of basis operators is comparable to the number
of contributing states. Our operator set is too small to use this
method effectively but it is of interest nonetheless to compare this
method with the results of the multiexponential fit.
\begin{figure}
\scalebox{0.50}{\includegraphics*{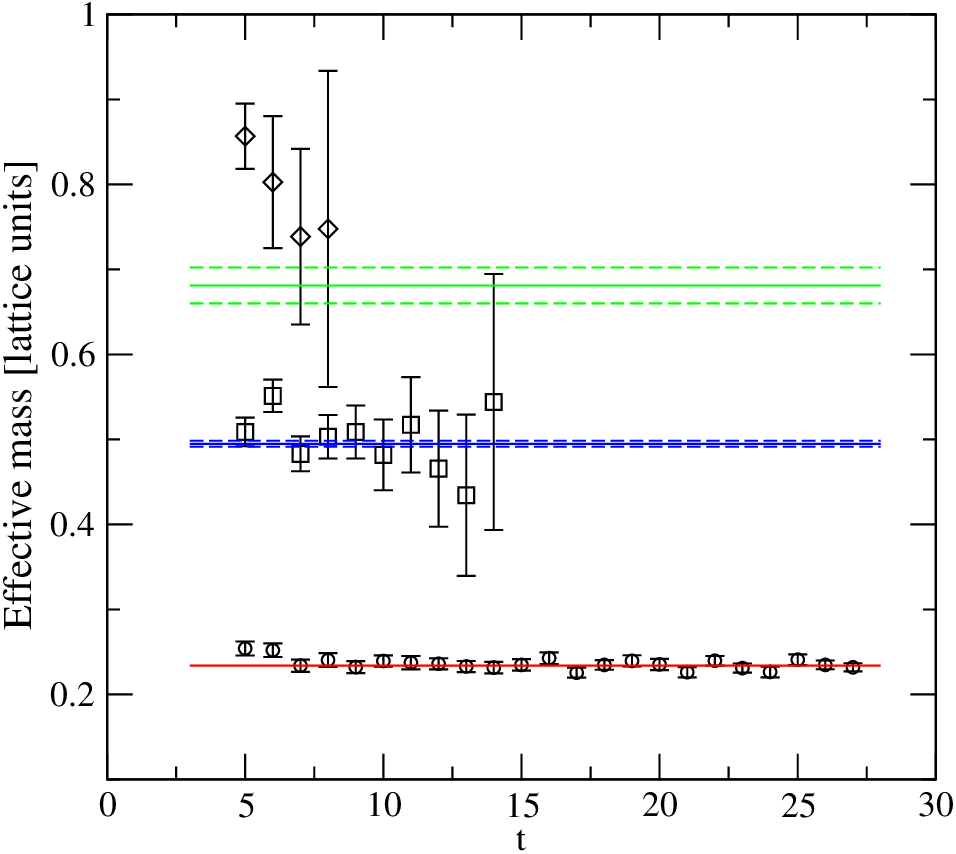}}
\caption{The effective mass of the eigenvalues from a variational analysis of the
correlator matrix for the pseudoscalar meson. The lines show the simulation energies
for the lowest three states from the 10-term multiexponential fit.}
\label{psgevp}
\end{figure}
\begin{figure}
\scalebox{0.50}{\includegraphics*{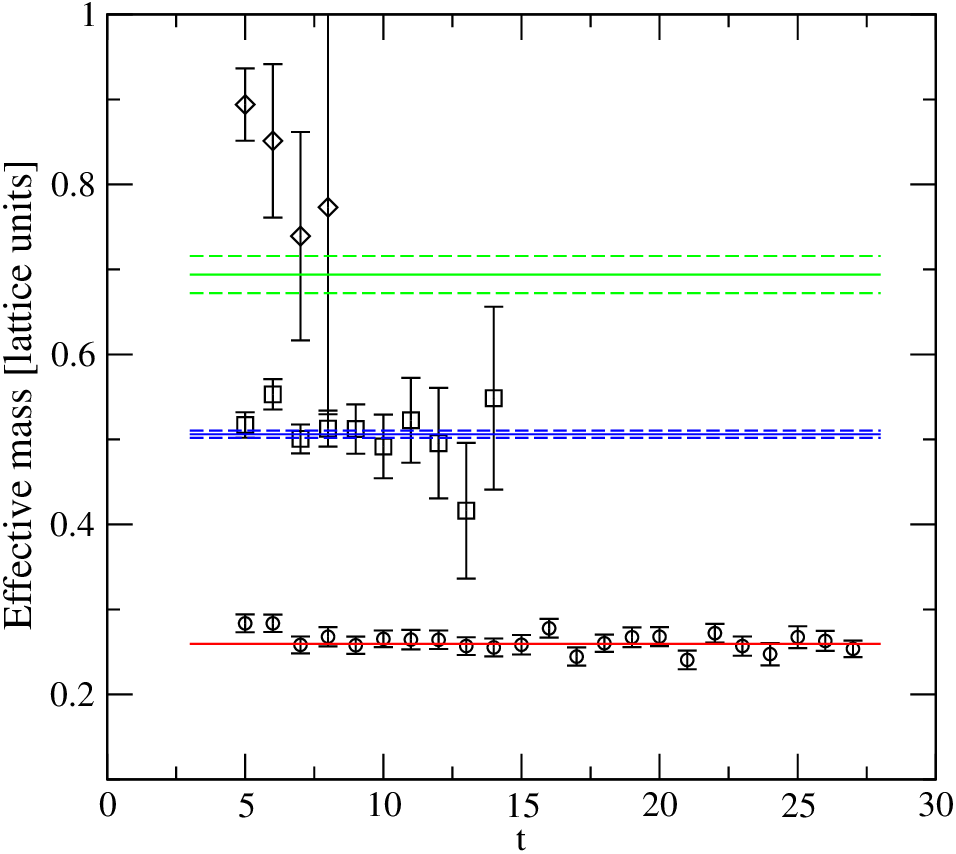}}
\caption{The effective mass of the eigenvalues from a variational analysis of the
correlator matrix for the vector meson. The lines show the simulation energies
for the lowest three states from the 10-term multiexponential fit.}
\label{vecgevp}
\end{figure}
Figs.~\ref{psgevp} and \ref{vecgevp} show
the effective mass plots for the eigenvalues $\lambda_{k}(t)$ which
are solutions of the generalized eigenvalue problem (with $t_0=4$)
\begin{equation}
g(t)f_{k}(t)=\lambda_{k}(t)g(t_{0})f_{k}(t)
\end{equation}
 where $f_{k}(t)$ is the eigenvector. For higher states only a limited
number of time steps are available before the effective mass degenerates
into noise. The lines on the plots show the simulation energies from
the 10-term multiexponential fit. For the ground and first-excited
states, where a meaningful comparison is possible, there is complete
consistency. 

The overlap coefficients in the vector channel provide another test
of the calculation. They can be used to determine the partial width
for Upsilon states to decay into lepton pairs and this can be compared
to experimental values. The decay width can be expressed in terms
of the wavefunction at the origin $\Psi_{n}(0)$ as (see \cite{gray})
\begin{equation}
\Gamma(\Upsilon(nS)\rightarrow e^{+}e^{-})=\frac{16}{9}\pi\alpha\frac{\left|\Psi_{n}(0)\right|^{2}}{M_{\Upsilon(nS)}^{2}}Z_{match}^{2}
\end{equation}
 where $\alpha$ is the electromagnetic coupling constant and the
matching factor $Z_{match}$ relates the lattice vector current to
the renormalized continuum current.
 With nonrelativistic normalization
of states, $\Psi(0)$ is related to the overlap coefficient of the local
vector operator by $\Psi(0)=c_{l}/\sqrt{6}.$ The results from our
calculation are shown in Table \ref{tab_leptonic}. The matching factor $Z_{match}$ 
has not been calculated for the version of lattice NRQCD that we use so
the leading order value of 1 has been assumed. 
Hart \emph{et al.}\cite{hart07} have computed that
matching coefficient for lattice NRQCD with stability parameter equal
2. Using their result as a guide we might expect effects from matching
of about 5\% but without an actual calculation it is not possible
to say with certainty which way they would go. Given this state of
the calculation, consistency with experiment is reasonable.

\begin{table}

\caption{Decay amplitude and partial width for Upsilon leptonic decay. The
experimental values $\Gamma_{exp}$ are from the Particle Data Group\cite{pdg10}.}

\begin{centering}
\begin{tabular}{cccc}
\hline 
State&
$a^{3/2}\Psi(0)$&
$\Gamma${[}keV]&
$\Gamma_{exp}${[}keV]\tabularnewline
\hline
\hline 
$\Upsilon$(1S)&
0.18418(7)&
1.16(2)&
1.34(2)\tabularnewline
$\Upsilon$(2S)&
0.1397(33)&
0.595(28)&
0.612(11)\tabularnewline
$\Upsilon$(3S)&
0.156(24)&
0.70(21)&
0.443(8)\tabularnewline
\hline
\end{tabular}
\par\end{centering}
\label{tab_leptonic}
\end{table}

\section{\label{sec_threepnt}Three-point functions}

Three-point functions describing the vector to pseudoscalar transition
induced by a current operator insertion are constructed using a sequential 
source method. The transition operator is taken here to be 
just the leading nonrelativistic operator $\sigma$ which acting on a 
quark (or antiquark) converts a vector state to a pseudoscalar 
(or vice versa). Starting with a vector (pseudoscalar)
source at $t_{s}$ the quark propagator is evolved to some maximum
time $T$ at which a pseudoscalar (vector) operator is applied. This
quantity is then evolved backward in time. At intermediate times $t_{s}<t'<T$
the current operator is inserted and evolution is continued to complete
the quark antiquark loop at the source. Appropriate momentum projections
are applied at the source, sink and current insertion to ensure momentum
conservation. The vector meson is always projected to have zero momentum;
the pseudoscalar recoils against the momentum carried by the current.

With a vector operator at the source and pseudoscalar at the sink
the three-point function is expected to have the form
\begin{eqnarray}\label{eq_threepnt}
\nonumber
G_{oo'}^{(VP)}(t';T) &=& \sum_{n,n'}c_{o}^{(V)}(n)A_{nn'}^{(VP)}c_{o'}^{(P)}(n') \\
                     && \times e^{-E_{n}^{(V)}(t'-t_{s})}e^{-E_{n'}^{(P)}(T-t')}
\end{eqnarray}
where the subscripts $o,o'$ indicate the type of smearing used ($l, s$ or $d$).
 The overlap coefficients and simulation energies are the same ones
that appear in the two-point function but now have a superscript attached
to distinguish between vector and pseudoscalar states. The quantity
$A_{nn'}^{(VP)}$ is the matrix element of the transition operator between
the vector state $n$ and the pseudoscalar state $n'$. This is identified
with the wavefunction overlap appearing in (\ref{eq_overlapint}). The three-point 
function with pseudoscalar source and vector sink has the same form with $V$
and $P$ labels reversed. The matrix elements $A_{nn'}^{(PV)}$ are related
to those appearing in (\ref{eq_threepnt}) by $A_{nn'}^{(PV)}=A_{n'n}^{(VP)}.$
The matrix elements can be determined by fitting the $t'$ dependence of the 
three-point function for a fixed $T$.

If the spin-dependent interaction terms in the NRQCD Hamiltonian,
which are nonleading in the nonrelativistic expansion, were omitted
and pseudoscalar meson recoil momentum set to zero, the three-point
functions would be independent of $t'$ and numerically equal to the
two-point functions at time separation $T-t_{s}$ for all $T$. The solution
for the matrix elements would be trivial $A_{nn'}^{(VP)}=\delta_{nn'},$
the same as for the wavefunction overlap (\ref{eq_overlapint}) in the extreme 
nonrelativistic limit\cite{bram06}.

Three-point functions were computed for three values of pseudoscalar
recoil momentum corresponding to (0,0,0), (1,0,0) and (2,0,0) and
for two source-sink time separations $T-t_{s}$ equal to 27 and 19.
All combinations of operator types l,s,d were calculated but in the
analysis only the combinations ll, ls, sl, ld, dl, ss were used. Fits
were done both including and excluding the ss operator combination.
The other combinations were noisy and not useful in pulling out the
excited to ground-state transitions that are of interest here. 

\begin{figure}
\scalebox{0.50}{\includegraphics*{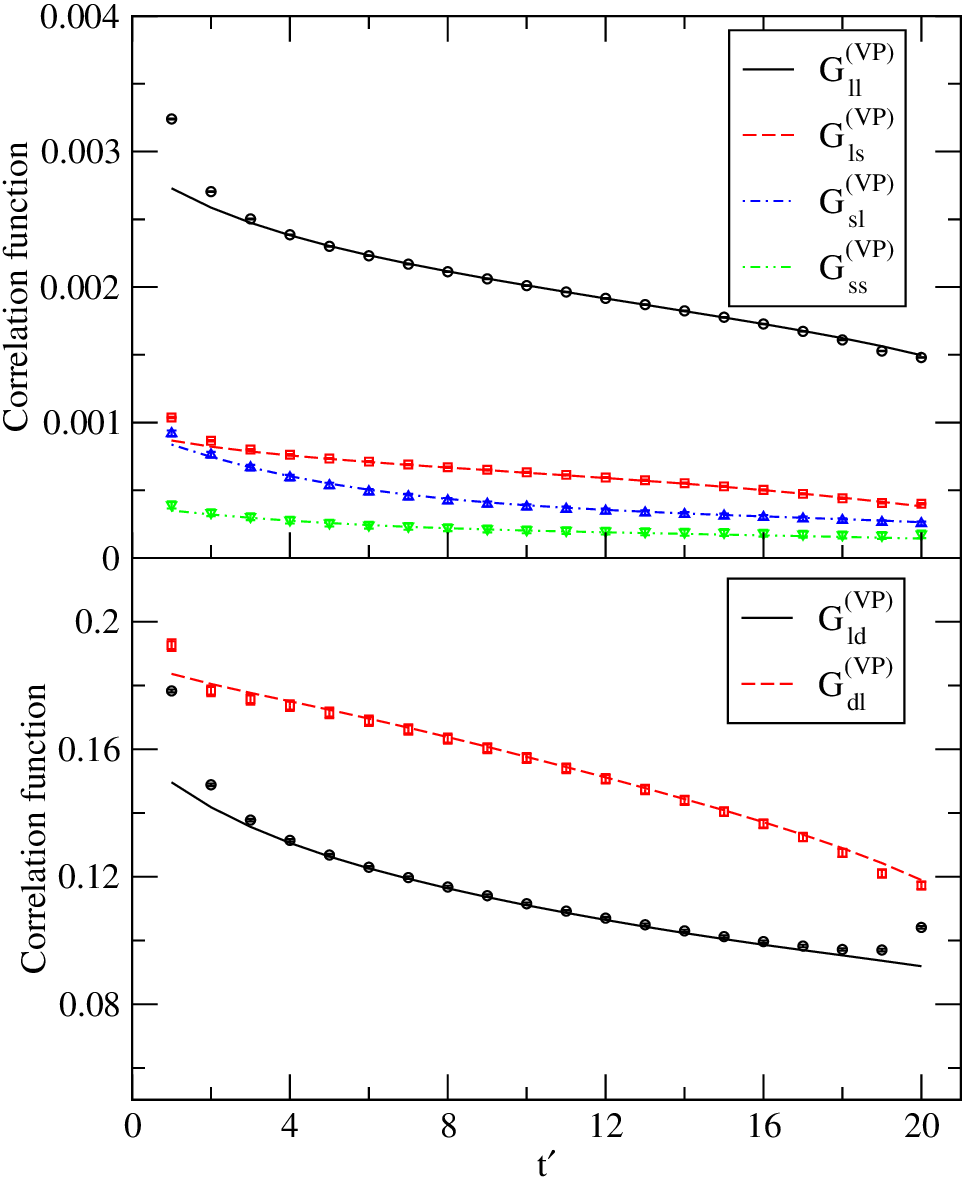}}
\caption{Three-point correlation functions with vector source and pseudoscalar
sink at time separation $T-t_{s}$ equals 19 for different operator combinations.
Symbols are simulation values and the lines are the result of a simultaneous
fit. The $t'$ fit range is 4 to 17.}
\label{threepntcfVP}
\end{figure}

\begin{figure}
\scalebox{0.50}{\includegraphics*{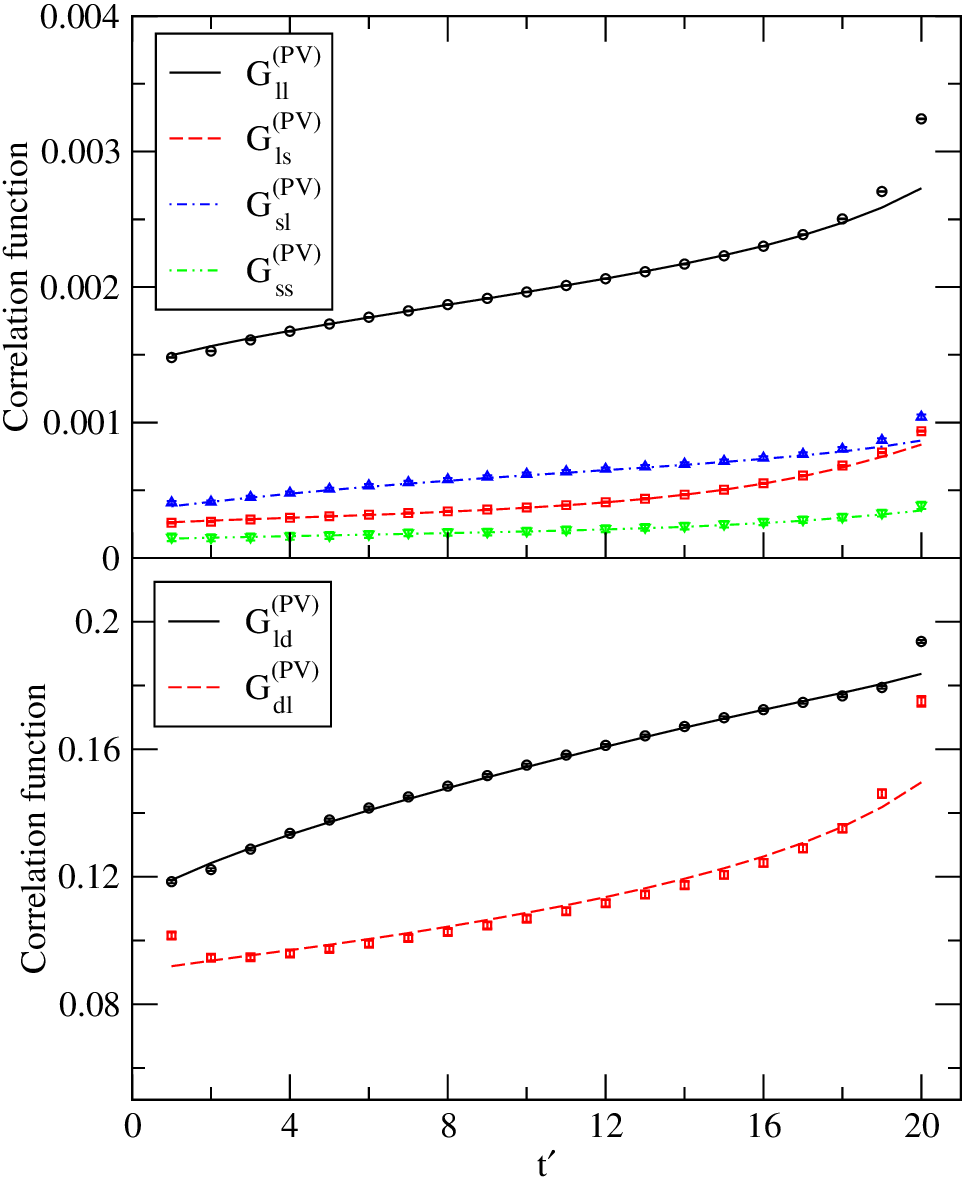}}
\caption{Three-point correlation functions with pseudoscalar source and vector
sink at time separation $T-t_{s}$ equals 19 for different operator combinations.
Symbols are simulation values and the lines are the result of a simultaneous
fit. The $t'$ fit range is 4 to 17.}
\label{threepntcfPV}
\end{figure}

\begin{figure}
\scalebox{0.50}{\includegraphics*{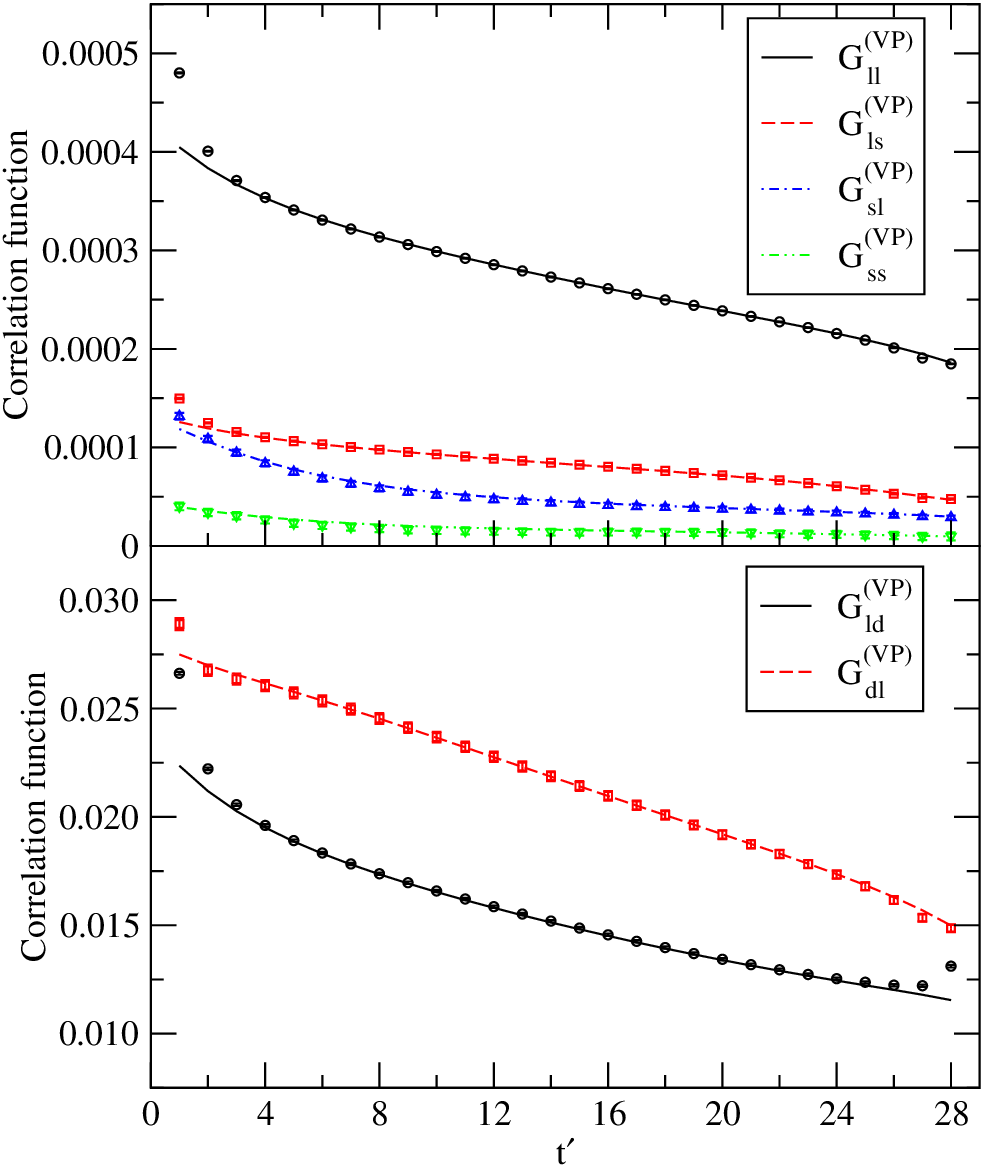}}
\caption{Three-point correlation functions with vector source and pseudoscalar
sink at time separation $T-t_{s}$ equals 27 for different operator combinations.
Symbols are simulation values and the lines are the result of a simultaneous
fit. The $t'$ fit range is 4 to 25.}
\label{threepntcfVP28}
\end{figure}

\begin{figure}
\scalebox{0.50}{\includegraphics*{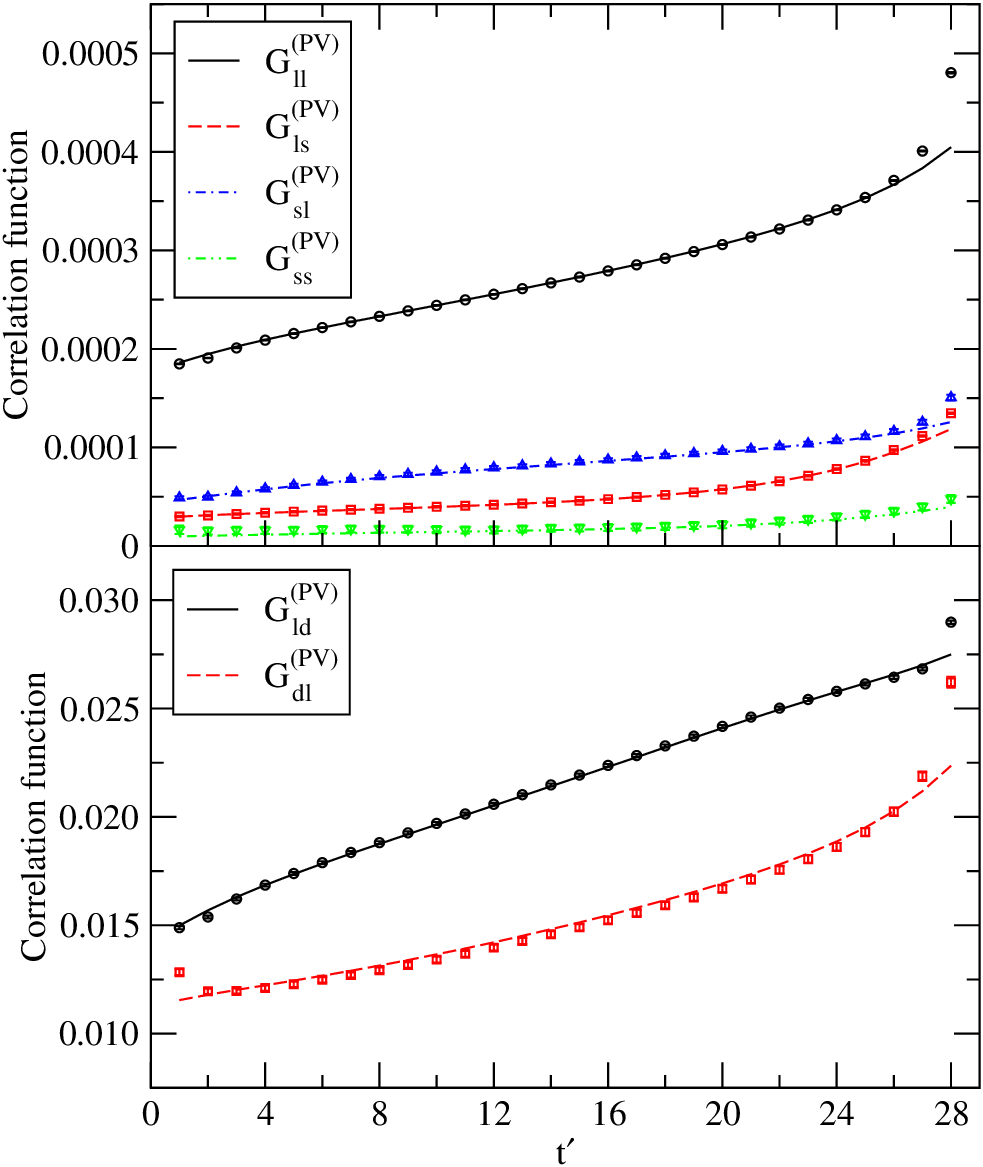}}
\caption{Three-point correlation functions with pseudoscalar source and vector
sink at time separation $T-t_{s}$ equals 27 for different operator combinations.
Symbols are simulation values and the lines are the result of a simultaneous
fit. The $t'$ fit range is 4 to 25.}
\label{threepntcfPV28}
\end{figure}

The two-point functions were fit with many exponential terms in order
to get a stable result for the lowest few states. For the three-point
function with large source-sink separation the contribution of the
high-lying states is highly suppressed and neglected in the fit of
the $t'$ dependence. Only the lowest three states were considered and
the $nn'$ combinations included in the sums in (\ref{eq_threepnt}) 
were 11, 12, 21, 13, 31, 22. As a test of the robustness of the results 
some five-parameter
fits, excluding the 22 term, were done. The time fit range was taken
to be $t_{s}+2<t'<T-2.$ The fits using two-point function parameters,
overlap coefficients and simulation energies, determined using ten
terms are given here. Using eight-term two-point function parameters
gives essentially the same values. The six-term two-point function
parameters lead to slightly different results but we do not consider
six terms to be sufficient for the two-point function fit. The determination
of the matrix elements is done by a simultaneous fit to a set of three-point
functions. 
Statistical errors are estimated using a bootstrap analysis. 
Some representative three-point correlation functions and fits are given 
in Figs.~\ref{threepntcfVP} and \ref{threepntcfPV} for $T-t_{s}$ equal to 19
and in Figs.~\ref{threepntcfVP28} and \ref{threepntcfPV28} for $T-t_{s}$ equal to 27.

The results for the three-point matrix elements are given in Tables
\ref{tab_mom0} - \ref{tab_mom2} for different values of recoil momentum. For $T-t_{s}=19$ the
2 to 2 transition is clearly necessary to get results that are consistent
with the larger time separation. For the excited to ground-state transitions,
that are of primary interest, there is very good agreement between
results using the shorter and longer time separation.

\begin{table}

\caption{Three-point matrix elements from simultaneous fits to $N_{cf}$ correlation
functions and with different numbers of parameters at zero recoil momentum.}

\begin{centering}
\begin{tabular}{ccccccc}
\hline 
$N_{cf}$&
$A_{11}^{(VP)}$&
$A_{21}^{(PV)}$&
$A_{31}^{(PV)}$&
$A_{21}^{(VP)}$&
$A_{31}^{(VP)}$&
$A_{22}^{(VP)}$\tabularnewline
\hline
\hline 
&
&
\multicolumn{3}{c}{$T-t_{s}=19$}
&
&
\tabularnewline
10&
0.916(2)&
-0.043(7)&
-0.069(6)&
0.090(7)&
0.052(5)&
\tabularnewline
10&
0.915(2)&
-0.068(2)&
-0.050(4)&
0.072(4)&
0.065(3)&
1.11(31)\tabularnewline
12&
0.915(2)&
-0.068(3)&
-0.050(4)&
0.071(4)&
0.065(3)&
1.11(23)\tabularnewline
&
&
\multicolumn{3}{c}{$T-t_{s}=27$}
&
&
\tabularnewline
10&
0.916(2)&
-0.062(7)&
-0.056(7)&
0.075(7)&
0.059(6)&
\tabularnewline
10&
0.916(2)&
-0.068(3)&
-0.050(6)&
0.071(3)&
0.062(4)&
2.1(2.2)\tabularnewline
12&
0.916(2)&
-0.068(3)&
-0.051(6)&
0.071(4)&
0.062(4)&
1.9(1.8)\tabularnewline
\hline
\end{tabular}
\par\end{centering}
\label{tab_mom0}
\end{table}

\begin{table}

\caption{Three-point matrix elements from simultaneous fits to $N_{cf}$ correlation
functions and with different numbers of parameters at one unit of recoil momentum.}

\begin{centering}
\begin{tabular}{ccccccc}
\hline 
$N_{cf}$&
$A_{11}^{(VP)}$&
$A_{21}^{(PV)}$&
$A_{31}^{(PV)}$&
$A_{21}^{(VP)}$&
$A_{31}^{(VP)}$&
$A_{22}^{(VP)}$\tabularnewline
\hline
\hline 
&
&
\multicolumn{3}{c}{$T-t_{s}=19$}
&
&
\tabularnewline
10&
0.908(1)&
-0.042(8)&
-0.060(8)&
0.095(7)&
0.057(5)&
\tabularnewline
10&
0.907(1)&
-0.062(6)&
-0.047(7)&
0.079(4)&
0.068(5)&
0.92(27)\tabularnewline
12&
0.907(1)&
-0.062(6)&
-0.047(7)&
0.079(5)&
0.067(5)&
0.95(21)\tabularnewline
&
&
\multicolumn{3}{c}{$T-t_{s}=27$}
&
&
\tabularnewline
10&
0.908(2)&
0.057(8)&
-0.052(9)&
0.082(6)&
0.063(6)&
\tabularnewline
10&
0.907(2)&
0.061(5)&
-0.048(8)&
0.079(4)&
0.066(6)&
1.6(1.9)\tabularnewline
12&
0.907(2)&
0.061(5)&
-0.048(8)&
0.079(5)&
0.066(6)&
1.6(1.5)\tabularnewline
\hline
\end{tabular}
\par\end{centering}
\label{tab_mom1}
\end{table}

\begin{table}

\caption{Three-point matrix elements from simultaneous fits to $N_{cf}$ correlation
functions and with different numbers of parameters at two units of recoil momentum.}

\begin{centering}
\begin{tabular}{ccccccc}
\hline 
$N_{cf}$&
$A_{11}^{(VP)}$&
$A_{21}^{(PV)}$&
$A_{31}^{(PV)}$&
$A_{21}^{(VP)}$&
$A_{31}^{(VP)}$&
$A_{22}^{(VP)}$\tabularnewline
\hline
\hline 
&
&
\multicolumn{3}{c}{$T-t_{s}=19$}
&
&
\tabularnewline
10&
0.878(1)&
-0.010(6)&
-0.055(6)&
0.116(7)&
0.066(6)&
\tabularnewline
10&
0.877(1)&
-0.030(4)&
-0.041(6)&
0.101(5)&
0.078(6)&
1.01(25)\tabularnewline
12&
0.877(1)&
-0.031(4)&
-0.041(6)&
0.102(5)&
0.078(6)&
1.02(20)\tabularnewline
&
&
\multicolumn{3}{c}{$T-t_{s}=27$}
&
&
\tabularnewline
10&
0.878(1)&
-0.026(6)&
-0.041(8)&
0.104(6)&
0.066(8)&
\tabularnewline
10&
0.878(2)&
-0.031(4)&
-0.037(8)&
0.101(5)&
0.070(6)&
1.9(1.8)\tabularnewline
12&
0.878(2)&
-0.029(5)&
-0.039(7)&
0.100(5)&
0.068(6)&
1.0(1.6)\tabularnewline
\hline
\end{tabular}
\par\end{centering}
\label{tab_mom2}
\end{table}

\begin{figure}
\scalebox{0.50}{\includegraphics*{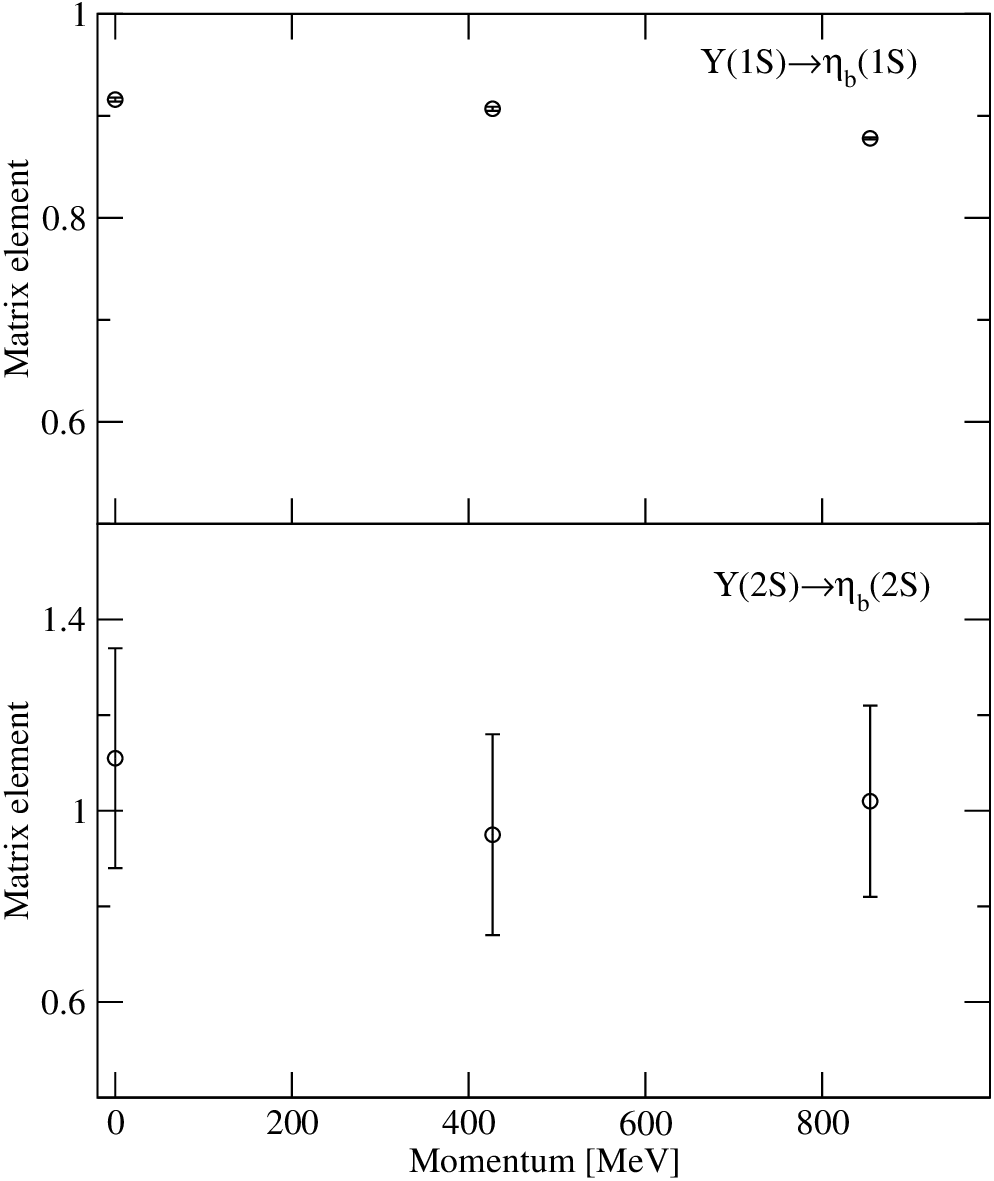}}
\caption{The matrix elements for decay of $\Upsilon$ to $\eta_{b}$ with the same
principal quantum number as a function of momentum.}
\label{M1momupsg}
\end{figure}

\begin{figure}
\scalebox{0.50}{\includegraphics*{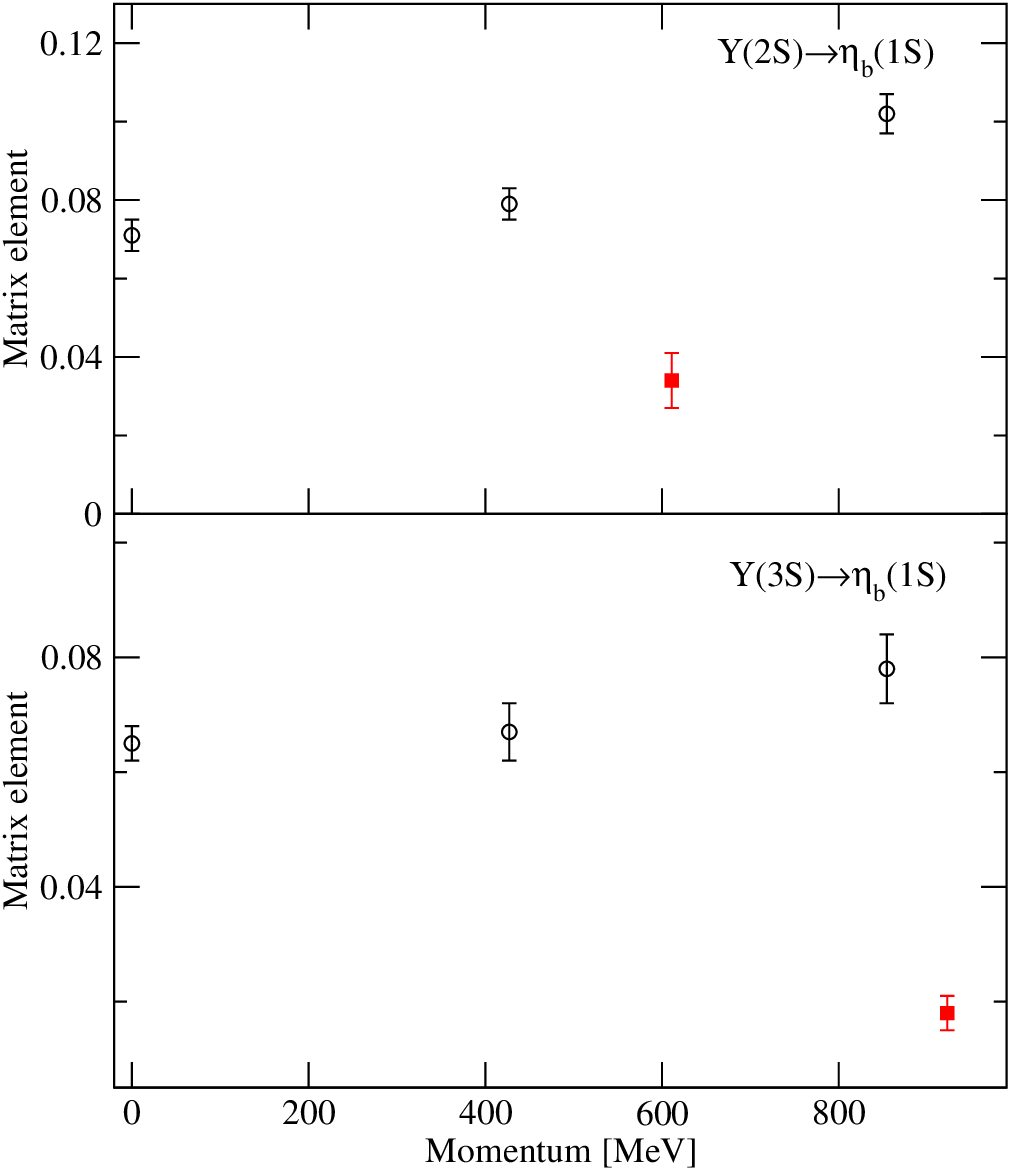}}
\caption{The matrix elements for decay of an excited $\Upsilon$ to the $\eta_{b}$ 
ground state as a function of momentum. The square symbols show the matrix element
values inferred from the measured decay widths\cite{babar08,babar09}.}
\label{M1momups}
\end{figure}

\begin{figure}
\scalebox{0.50}{\includegraphics*{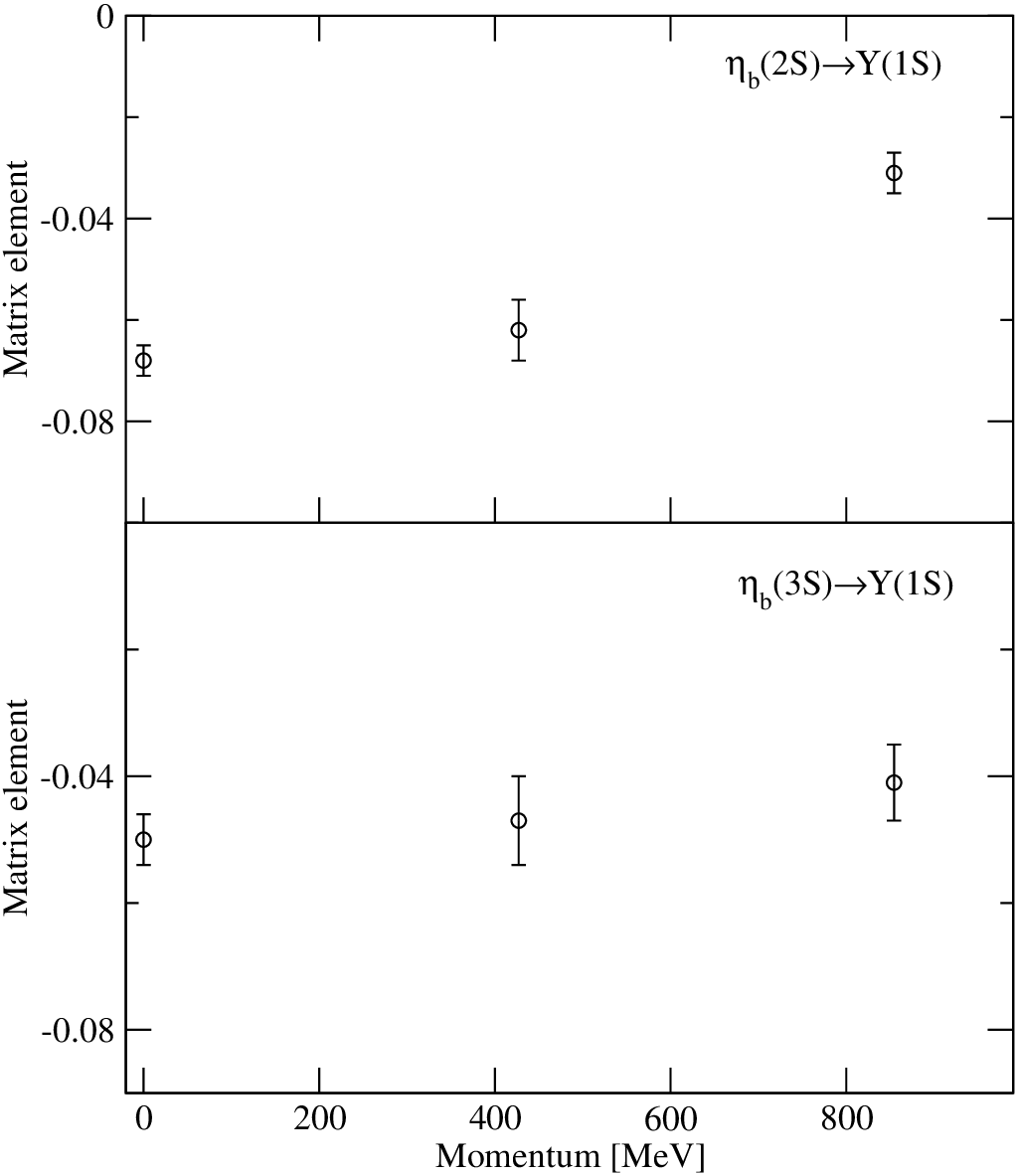}}
\caption{The matrix elements for decay of an excited $\eta_{b}$ to the $\Upsilon$  
ground state as a function of momentum.}
\label{M1mometa}
\end{figure}

The results from the $T-t_{s}=19$ analysis with a six parameter fit
are plotted in Fig.~\ref{M1momupsg} to \ref{M1mometa} as a function of momentum. 
In Fig.~\ref{M1momups} the matrix
elements inferred from the measured $\Upsilon(2S)$ and $\Upsilon(3S)$
to $\eta_{b}$ partial widths \cite{babar08,babar09} 
are also shown at the physical momentum
for these decays. The features of the results are easily understood.
The matrix elements for the $\Upsilon(1S)$ to $\eta_{b}(1S)$ and
$\Upsilon(2S)$ to $\eta_{b}(2S)$ transitions
are close to 1 since the wavefunctions
of the states involved are very similar. The $\Upsilon(1S)$ to $\eta_{b}(1S)$
matrix element decreases only very slowly with recoil momentum which
reflects the small size of bottomonium. The excited state to ground-state
transitions have matrix elements that are small in magnitude
due to near orthogonality of wavefunctions. The relative negative
sign of $\Upsilon\rightarrow\eta_{b}$
and $\eta_{b}\rightarrow\Upsilon$ reflects the fact that the dominant
spin-dependent quark antiquark interaction acts with different sign
in pseudoscalar and vector states. The recoil effect contributes positively
to all transitions (see, for example, \cite{bram06}) which explains the 
momentum dependence.

The calculated matrix elements are large compared to the empirical
values inferred from the measured partial widths. However, there are
a variety of improvements that are needed before definitive conclusions
can be drawn. The lattice vector current operator has to be matched
to the renormalized continuum current as discussed in connection to
Upsilon leptonic decay\cite{gray,hart07}.  Relativistic corrections
are not incorporated into the transition operator and these are likely 
to be important for the hindered excited state decays\cite{bram06}. 
As well, one might ask about $\mathcal{O}(v^{6})$ terms\cite{meinel10} 
and radiative corrections (beyond tadpole improvement)
to spin-dependent interactions in the Hamiltonian which have
been shown to have a noticeable effect on the $\Upsilon-\eta_{b}$
mass splitting\cite{hamm11}. Finally, there is the question of continuum 
extrapolation which this calculation, done at a single lattice spacing, 
can not address.

There are other systematics that we can not deal with quantitatively
but which it is reasonable to think are small. Bottomonium has only
heavy valence quarks so extrapolation to the physical point for up
and down quarks comes in only through the influence of sea quarks
on the gauge field. Since the simulation is done very near the physical
point, with quarks which give a pion mass of 156MeV, it would be surprising
if a simulation at the physical point would be much different. Finite
volume can also lead to a significant systematic effect in lattice
simulations but is unlikely to be the case here. The size of bottomonium
is much smaller than the spatial lattice size so finite volume effects
arise indirectly through light quarks in the sea. We do not have any
estimates of this effect. However, finite volume effects have been
studied for heavy-light mesons\cite{colan10}.
For our lattice size they are very small and it is reasonable 
to expect them to be even smaller for bottomonium.

\section{\label{sec_summ}Summary}

Bottomonium S-wave states were studied using lattice NRQCD focusing on the
low-lying excited states. It was found that using a set of operators, 
including smeared operators which suppress the ground-state contribution to
the correlation functions, robust results for the lowest few states could
be obtained. Constrained multiexponential fitting\cite{constrainedfit} was used
for the analysis of two-point correlation functions. As a check, an analysis 
based on the variational method\cite{luscher90,michael85} was carried out. 
Where a meaningful comparison could be made, the two analysis methods 
gave consistent results.

Mass differences between the Upsilon ground state and the first two
excited states are in reasonable agreement with experimental values.
The mass difference between $\Upsilon$ and $\eta_{b}$ is not well
reproduced by the calculation. Issues such as continuum extrapolation,
higher order nonrelativistic terms and radiative corrections to the
NRQCD Hamiltonian have to be considered.

A primary goal of this study was to see if the highly suppressed
matrix elements of excited state transitions could be extracted. 
Three-point functions for transitions from vector to pseudoscalar states
with the leading nonrelativistic M1 operator were calculated. 
Using overlap coefficients and simulation energies obtained from fitting 
the two-point functions, a simultaneous fit was done to sets 
of three-point functions. Transition matrix elements with reasonably
small statistical errors could be obtained for a number of excited
state decays. The results were very stable with respect to choice 
of two-point function parameters and the number of three-point functions 
and matrix elements included in the fit.

The qualitative features of the calculated matrix elements are
as expected. The matrix elements for the $\Upsilon(1S)$ to $\eta_{b}(1S)$ 
and $\Upsilon(2S)$ to $\eta_{b}(2S)$ transitions are close to one.
For states identified with different principal quantum numbers the
transitions are highly suppressed. The relative negative
sign of $\Upsilon\rightarrow\eta_{b}$ and $\eta_{b}\rightarrow\Upsilon$
matrix elements can be understood by considering perturbatively the effect 
of spin-dependent interactions. The qualitative momentum dependence 
is in accord with, for example, pNRQCD\cite{bram06}.

Quantitatively the values obtained here for excited $\Upsilon$
to ground-state $\eta_{b}$ matrix elements are considerably
larger than the values inferred from the experimentally determined
decay widths. These decays are dependent on the interplay of small
effects and are likely to be sensitive to relativistic corrections 
to the transition operator. As well,
the issues that enter into the $\Upsilon$ - $\eta_{b}$
spin splitting, e.g., relativistic corrections to the NRQCD 
Hamiltonian\cite{meinel10}, operator matching\cite{hamm11} and 
continuum extrapolation\cite{gray} have to be dealt with to get 
definitive results.

\acknowledgments
We thank the PACS-CS Collaboration for making their dynamical gauge 
field configurations available. As well, we thank G. von Hippel, 
J. Shigemitsu and H. Trottier for some helpful comments.
This work was supported in part by the 
Natural Sciences and Engineering Research Council of Canada.



\begin{thebibliography}{99}

\bibitem{herb77}
S.~W.~Herb {\emph {et al}}.,
Phys. Rev. Lett. \textbf{39}, 252 (1977).

\bibitem{innes77}
W.~R.~Innes {\emph {et al}}.,
Phys. Rev. Lett. \textbf{39}, 1240 (1977).

\bibitem{babar08}
B.~Aubert {\emph {et al}}. (BABAR Collaboration),
Phys. Rev. Lett. \textbf{101}, 071801 (2008).

\bibitem{babar09}
B.~Aubert {\emph {et al}}. (BABAR Collaboration),
Phys. Rev. Lett. \textbf{103}, 161801 (2009).

\bibitem{latnrqcd}
G.~P.~Lepage, L.~Magnea, C.~Nakhleh, U.~Magnea, and K.~Hornbostel,
Phys. Rev. D \textbf{46}, 4052 (1992).

\bibitem{godf01}
S.~Godfrey and J.~L.~Rosner,
Phys. Rev. D \textbf{64}, 074011 (2001).
	
\bibitem{bram06}
N.~Brambilla, Y.~Jia and A.~Vairo,
Phys. Rev. D \textbf{73}, 054005 (2006).
	
\bibitem{ebert03}
D.~Ebert, R.~N.~Faustov and V.~O.~Galkin,
Phys. Rev. D \textbf{67}, 014027 (2003).

\bibitem{rmw86}
R.~M.~Woloshyn,
Z. Phys. C \textbf{33}, 121, (1986).

\bibitem{cris92}
M.~Crisafulli and V.~Lubicz,
Phys. Lett. B \textbf{278}, 323 (1992).

\bibitem{dudek06}
J.~J.~Dudek, R.~G.~Edwards and D.~G.~Richards,
Phys. Rev. D \textbf{73}, 074507 (2006).

\bibitem{dudek09}
J.~J.~Dudek, R.~G.~Edwards and C.~E.~Thomas,
Phys. Rev. D \textbf{79}, 094504 (2009).

\bibitem{chen11}
Y.~Chen {\emph {et al}.}, arXiv:1104.2655 [hep-lat].

\bibitem{luscher90}
M.~L\"uscher and U. Wolff,
Nucl. Phys. \textbf{B339}, 222 (1990).

\bibitem{michael85}
C.~Michael,
Nucl. Phys. \textbf{B259}, 58 (1985).

\bibitem{constrainedfit}
G.~P.~Lepage  {\emph {et al}.},
Nucl. Phys. B (Proc.Suppl.) \textbf{106}, 12 (2002).

\bibitem{davies10}
C.~T.~H.~Davies, E.~Follana, I.~D.~Kendall, G.~P.~Lepage, and C.~McNeile,
Phys. Rev. D \textbf{81}, 034506 (2010).

\bibitem{pacscs09}
S.~Aoki {\emph {et al}}. (PACS-CS Collaboration),
Phys. Rev. D \textbf{79}, 034503 (2009).

\bibitem{davies94}
C.~T.~H.~Davies, K.~Hornbostel, A.~Langnau, G.~P.~Lepage, A.~Lidsey, J.~Shigemitsu, 
and J.~Sloan,
Phys. Rev. D \textbf{50}, 6963 (1994).

\bibitem{gray}
A.~Gray, I.~Allison, C.~T.~H.~Davies, E.~Gulez, G.~P.~Lepage, J.~Shigemitsu, and M.~Wingate,
Phys. Rev. D \textbf{72}, 094507 (2005).

\bibitem{lewis09}
R.~Lewis and R.~M.~Woloshyn,
Phys. Rev. D \textbf{79}, 014502 (2009).

\bibitem{gusken89}
S. G\"usken, U.~L\"ow, K.-H.~M\"utter, R.~Sommer, A.~Patel, and K.~Schilling, 
Phy. Lett. B \textbf{227}, 266 (1989).

\bibitem{davies07}
C.~T.~H.~Davies, E.~Follana, K.~Y.~Wong, G.~P.~Lepage, J.~Shigemitsu,
PoS LAT2007, 378 (2007).

\bibitem{pdg10}
K. Nakamura {\emph {et al}}. (Particle Data Group), 
J. Phys. G \textbf{37}, 075021 (2010). 

\bibitem{meinel09}
S.~Meinel,
Phys. Rev. D \textbf{79}, 094501 (2009).

\bibitem{cleo10}
G.~Bonvicini {\emph {et al}}. (CLEO Collaboration),
Phys. Rev. D \textbf{81}, 031104 (2010).

\bibitem{meinel10}
S.~Meinel,
Phys. Rev. D \textbf{82}, 114502 (2010).

\bibitem{hamm11}
T.~C.~Hammant, A.~G.~Hart, G.~M.~von~Hippel, R.~R.~Horgan and C.~J.~Monahan, 
arXiv:1105.5309 [hep-lat].

\bibitem{bloss09}
B.~Blossier, M.~Della Morte, G.~von Hippel, T.~Mendes and R.~Sommer,
JHEP \textbf{04}, 094 (2009).

\bibitem{hart07}
A.~Hart, G.~M.~von~Hippel and R.~R.~Horgan,
Phys. Rev. D \textbf{75}, 014008 (2007).
 
\bibitem{colan10}
G.~Colangelo, A.~Fuhrer and S.~Lanz,
Phys. Rev. D \textbf{82}, 034506 (2010).
 
\end{thebibliography}
\end{document}